\documentclass[final,5p,times,twocolumn]{elsarticle}
\journal{Computer Physics Communications}
\usepackage{graphicx}
\usepackage[dvipsnames]{xcolor}
\usepackage{tikz}
\usetikzlibrary{decorations.pathmorphing}
\usetikzlibrary{arrows}
\usepackage{amssymb}
\usepackage{amsmath}
\usepackage{booktabs}
\biboptions{sort&compress}
\usepackage{hyperref}
\usepackage{xspace}
\newcommand{\spec}{\textit{SpectraMatcher}\xspace}
\newcommand{\specs}{\textit{SpectraMatcher}'s\xspace}

\begin{document}
\begin{frontmatter}

\title{SpectraMatcher: A Python Program for\\
Interactive Analysis and Peak Assignment of Vibronic Spectra}

\author{Johanna Langner\textsuperscript{\dag}}
\author{Isabelle Weber}
\author{Henryk A. Witek\corref{hw}}
\author{Yuan-Pern Lee\corref{ypl}}

\cortext[hw]{\,Corresponding author. \textit{Email address:} hwitek@nycu.edu.tw}
\cortext[ypl]{\,Corresponding author. \textit{Email address:} yplee@nycu.edu.tw}

\address{Department of Applied Chemistry, National Yang Ming Chiao Tung University,
1001 University Road, Hsinchu, 300, Taiwan}

\begin{abstract}
\textit{SpectraMatcher} is a cross-platform desktop application for interactive comparison of experimental and computed vibronic spectra, designed to assist in the recognition and assignment of spectral patterns. It provides an intuitive graphical interface --- with no coding or scripting required --- for importing experimental spectra, visualizing them alongside the corresponding theoretical spectra constructed from Gaussian frequency calculations, and adjusting key parameters such as peak width, intensity scaling factors, and vibration-type-specific anharmonic corrections. \textit{SpectraMatcher} features an automated peak-matching algorithm that assigns experimental and computed peaks based on their intensity ratio and proximity. Assignments and spectra can be exported in multiple formats for publication or for further analysis. The software remains responsive even for large datasets, and supports efficient and reproducible interpretation of vibronic spectra.\\

\noindent \textbf{PROGRAM SUMMARY}

\begin{small}
\noindent
{\em Program Title:} SpectraMatcher \\
{\em CPC Library link to program files:} (to be added by Technical Editor) \\
{\em Developer's repository link:} \href{https://github.com/giogina/SpectraMatcher}{https://github.com/giogina/SpectraMatcher} \\
{\em Licensing provisions:} MIT  \\
{\em Programming language:}  Python                                  \\
{\em Supplementary material:} \\ Documentation and manual: \href{https://spectramatcher.gitbook.io/spectramatcher }{https://spectramatcher.gitbook.io/spectramatcher }\\ Example files: \href{https://github.com/giogina/SpectraMatcher/demo}{https://github.com/giogina/SpectraMatcher/demo} \\
{\em Nature of problem:} 
Constructing theoretical vibronic spectra from quantum chemical calculations and comparing them with the corresponding experimental  spectra is one of the central tasks in computational vibrational spectroscopy. It often requires manual file handling, spectral alignment, peak identification, and correction factor tuning. This process can be time-consuming and prone to inconsistencies, especially for large polyatomic molecules or in the case of overlapping spectral features. Automating these steps in a reproducible and user-friendly way remains a significant challenge. In the current contribution we report \textit{SpectraMatcher}, a graphical environment for automatic construction of molecular vibronic spectra for polyatomic molecules from precomputed quantumchemical data, which addresses these challenges through interactive visualization, automated processing, and reproducible spectral analysis.\\
{\em Solution method:} 
\textit{SpectraMatcher} provides an  interactive graphical user interface (GUI) for importing Gaussian output files [1] containing vibronic emission and excitation transitions, visualizing the resulting theoretical vibronic spectra alongside their experimental counterparts, and applying real-time spectral adjustments to the computed data --- such as shifting, broadening and mode-type dependent anharmonic scaling corrections --- to improve agreement with experiment. The software uses \texttt{NumPy} [2] for efficient array-based convolution modeling of adjustable Lorentzian profiles. Automatic spectral feature detection is performed on experimental data via to facilitate the analysis, and an automated matching algorithm is designed to identify the corresponding theoretical peaks according to user-defined recognition thresholds. Resource-intensive operations are carried out via asynchronous task handling, ensuring that the GUI always remains responsive. The final results can be exported in publication-ready formats. The GUI is implemented using \texttt{DearPyGui} [3]. \\
{\em Additional comments including restrictions and unusual features:}\\
\textit{SpectraMatcher} currently only supports Gaussian 16 output files as the source of computed spectra. This limitation could readily be lifted by implementing additional parsers for other quantum chemistry software. The program stands out for its interactive design, along with features like type-specific anharmonic correction and visual peak assignment. It comes with extensive documentation and tutorials provided on GitBook.

\end{small}
\end{abstract}
\begin{keyword}
vibronic spectroscopy \sep peak matching \sep anharmonic correction \sep vibrational
frequency scale factors \sep convolution \sep spectrum analysis \sep fluorescence
\end{keyword}
\end{frontmatter}

\let\svthefootnote\thefootnote
\let\thefootnote\relax\footnotetext{
\hangindent=1.8em
\hangafter=1
\noindent\!\!\!\textsuperscript{\dag}\,\textit{For technical questions}: johanna.langner@nycu.edu.tw\\
\textit{For bugs or feature requests:} \href{https://github.com/giogina/SpectraMatcher/issues}{github.com/giogina/SpectraMatcher/issues}}
\addtocounter{footnote}{-1}\let\thefootnote\svthefootnote

\section{Introduction} \label{introduction}

Matching experimental and computed vibronic spectra --- i.e., spectra arising from molecular transitions altering simultaneously  electronic and vibrational quantum numbers --- is a crucial but labor-intensive task in theoretical spectroscopy. Traditionally, this process requires researchers to manually align spectral features, apply anharmonic corrections, and optimize parameters --- tasks that become increasingly challenging for complex spectra featuring hundreds of overlapping peaks. This is particularly evident in vibronic spectra of large conjugated systems \cite{hodecker2016, qian2020, plakhotnik2002, Crandall2020, zirkelbach2022, loe2024}, such as polycyclic aromatic hydrocarbons (PAHs) \cite{ruiterkamp2002, hoheisel2006, banasiewicz2007, makarewicz2012, Auerswald2013, Rice2015, Chakraborty2016, weberHydronaphtyl2022, weberSumanene2022, Katori2022, weberHexabenzacoronene2024, Schaefer2024}, where assigning peaks manually is a process that is time-consuming and prone to errors.

To address these challenges, we introduce in the current paper \spec, a desktop application that automates and streamlines the process of vibronic spectra matching. \spec has been created keeping in mind the needs of experimental groups working on the interpretation of recorded emission and excitation vibronic spectra. Built with Python \cite{van1995python, van2009python}, \spec integrates efficient NumPy-based vector operations \cite{numpy} required for spectra visualization and a responsive, intuitive graphical user interface (GUI) developed with DearPyGui \cite{dearpygui}. This interface provides real-time feedback for adjustments and corrections to the set of quantumchemical data used for the construction of theoretical spectra, facilitating their alignment with experimental observations. Standalone executable binaries are available for both Windows and Linux, allowing users to run the program without installing Python or executing scripts manually.

One of the most important features of \spec is the automatic deconvolution of the experimental band landscape in order to resolve it into individual peaks corresponding to transitions between specific vibronic levels of the studied molecule. Experimental spectra can be imported from common tabular formats such as plain text or spreadsheet files, with no special pre-processing required. Once loaded, candidate peak positions are detected as local maxima in the smoothed experimental data according to user-specified sensitivity criteria. Users can manually review and refine these  detected peaks to ensure their correctness in ambiguous or congested regions, or to add shoulders that escaped automatic detection. The extracted peak positions serve as anchor points for subsequent manipulations of the theoretical spectrum, enabling the assignment of the computed transitions, and efficient and reproducible interpretation of complex experimental data.

The theoretical spectra are constructed from Gaussian~16 \cite{g16} output files containing vibrational mode data and results of the Franck--Condon/Herzberg--Teller computations \cite{franck_elementary_1926, condon_theory_1926, herzberg_schwingungsstruktur_1933, barone2009}. Spectra for multiple excited states can be imported and overlaid simultaneously, with related Gaussian files automatically grouped based on the electronic state.
Theoretical spectra are displayed as both stick plots and convoluted profiles, allowing users to compare individual transitions with the overall spectral envelope. Mode labels can be shown in either Gaussian or Mulliken format; clicking on a label displays an animation of the corresponding vibrational mode.
These spectra can be interactively adjusted to improve agreement with experiment by tuning parameters such as peak half-width, zero-zero transition energy, and spectral intensity, to partially compensate systematic errors in theoretical calculations. Additionally, anharmonic correction factors can be applied separately to different types of vibrational modes, such as out-of-plane bends and X--H stretches \cite{pople_1981, scott_leastsquareScaling_1996, wheeler_2004, irikura_scaling_2009, jacobsen_notWorthwhile_2013}. The stick spectrum is color-coded to indicate the automatically detected mode type of each transition, clarifying which correction applies to which peaks.
All adjustments are immediately reflected in the spectral plots, allowing users to visually assess their effects and quickly converge toward a better match with the experimental data.

Once the spectra have been prepared, an adaptable peak-matching algorithm assigns theoretical transitions to their experimental counterparts. Matches are determined based on customizable thresholds for wavenumber proximity and relative intensity. The resulting assignments are displayed in a detailed table, which can be exported in CSV, Word, or LaTeX format for publication or further analysis.

By combining reproducible peak detection, interactive spectral adjustment, and automated peak matching, \spec enables efficient and consistent comparison between theoretical and experimental vibronic spectra.

This paper is structured as follows: The following Section~\ref{sec:vibronic_spectroscopy} provides a theoretical background on vibronic spectroscopy. Section~\ref{sec:workflow} outlines the program workflow, and Section~\ref{sec:implementation} discusses the software implementation. In Section~\ref{sec:case_study}, we demonstrate the use of \spec on a real-world example.

\section{Vibronic Spectroscopy}\label{sec:vibronic_spectroscopy}

This section provides a brief overview of the principles of vibronic spectroscopy. 
For a comprehensive introduction, the reader is referred to standard textbooks such as~\cite{herzberg_molecular_1950, banwell1994,atkins2002}.

\subsection{Vibronic Structure and Spectral Features}
When molecules absorb or emit light, they undergo transitions between distinct energy levels, usually labeled with different electronic, vibrational, and rotational quantum numbers. 
Electronic transitions are frequently accompanied by simultaneous changes in vibrational state (as well as, in the gas phase, rotational state). 
Combined \textit{vibr}ational and elect\textit{ronic} transitions are referred to as \textit{vibronic transitions}.

Electronic energy levels in molecules are typically separated by much larger energy gaps (above 10{,}000 $\text{cm}^{-1}$) than vibrational levels (below 4{,}000 $\text{cm}^{-1}$) \cite{banwell1994}. 
For each electronic state, the molecule can occupy many different vibrational energy levels, each corresponding to quantized excitations (indexed by a quantum number $v$) of different vibrational modes. Since non-radiative vibrational relaxation (energy redistribution) within an electronic state is generally fast, most vibronic transitions originate from the vibrational ground state ($v=0$) of the initial electronic state and proceed to various vibrationally-excited levels of the final electronic state ---  both in the cases of light \emph{absorption} (excitation from the  ground electronic state to an excited state) and \emph{emission} (radiative decay from an  excited electronic state back to the ground state).

As a result, at sufficient spectral resolution, an electronic molecular transition appears not as a single spectral band, but as a structured set of peaks corresponding to transitions to different vibrational levels of the final electronic state. The emerging spectral pattern is known as vibrational fine structure. In some cases, a single vibrational mode may manifest itself as a series of regularly spaced peaks, called a \textit{vibrational progression}. In larger molecules, however, such progressions are usually hidden between the multitude of overlapping transitions.

The lowest-energy vibronic transition, between the vibrational ground states of the initial and final electronic states, is called the \textit{zero-zero} transition. Its energy, denoted $\tilde{\nu}_{00}$, marks the onset of the vibronic band and serves as a reference point for locating other spectral features.

While the relative position, $\tilde{\nu} \!- \!\tilde{\nu}_{00}$, of a vibronic peak is often captured fairly accurately by quantum chemical methods, the absolute position of the vibronic band origin $\tilde{\nu}_{00}$ can be substantially offset due to limitations of the computational protocols used to determine electronic excitation energies and vibrational zero-point energies (ZPE) \cite{irikura_scaling_2009}. Consequently, computed spectra often need to be empirically shifted to align their $\tilde{\nu}_{00}$ with the experimental band origin before comparison. 

Occasionally, even the assignment of the correct excited state is ambiguous. In such cases, the application of \spec for a comparison of computed vibronic spectra for different excited states to the experimental line shape is a valuable asset to ensure correct assignment \cite{weber2024}.

Each vibronic transition can be characterized by the changes in vibrational quantum numbers $v$ of the involved normal modes. A transition from  $v\!=\!0$ to $v\!=\!1$ of a single vibrational mode is called a \textit{fundamental}, while transitions from  $v\!=\!0$ to $v\!=\!2, 3, \dots$ are referred to as \textit{overtones}. In systems with few active modes, overtones can produce the aforementioned vibrational progressions in the spectrum. When two or more vibrational modes are simultaneously excited, the resulting transitions are called \textit{combination bands}.

Vibronic transitions involving a single vibrational mode $m$ are denoted as $m_{v''}^{v'}$, where $v''$ and $v'$ are the vibrational quantum numbers in the electronic ground and excited state, respectively. For example, \( 10_0^1 \) denotes a fundamental transition of vibrational mode 10 from \( v'' = 0 \) to \( v' = 1 \) in an absorption process, while \( 10_2^0 \) represents an overtone transition from \( v' = 0 \) to \( v'' = 2 \) in an emission process.
Combination bands are represented by concatenating the signifiers of all involved modes (e.g., $4_0^110_0^112_0^2$).

\subsection{Computational Modeling of Vibronic Transitions}

\subsubsection{Franck--Condon principle and approximation}

\begin{figure}[tbh]
    \centering
    \scalebox{0.95}{\begin{tikzpicture}[scale=1.2]

\def\xmax{7}
\def\ymax{10}
\def\De{3}
\def\xground{2}
\def\yground{0.7}
\def\xexc{2.5}
\def\yexc{6}
\def\lambda{13}

\def\ntx{4}  
\def\ntm{1}  

\def\groundharmcolor{CornflowerBlue}
\def\groundmorsecolor{blue}
\def\excharmcolor{Thistle}
\def\excmorsecolor{Plum}

\draw[->] (0,0) -- (\xmax,0);
\node[below] at ({0.8*\xmax}, 0) {Nuclear displacement};
\draw[->] (0,0) -- (0,\ymax);
\node[above, rotate=90] at (0, \ymax-0.8) {Energy};

\begin{scope}
    
    \clip (0,0) -- (\xmax,0) -- (\xmax, \ymax) -- (0,\ymax) -- cycle;
    
    \begin{scope}
        \clip (0,\yexc-1.4) -- (\xmax,\yexc-1.4) -- (\xmax, 0) -- (0,0) -- cycle;
        \draw[domain=0:\xmax,smooth,variable=\x,\groundharmcolor,samples=50,thick] plot ({\x},{\yground+\De*(\x - \xground)^2});
        \draw[domain=0:\xmax,smooth,variable=\x,\groundmorsecolor,samples=50,thick] plot ({\x},{\yground + \De*(1 - exp(-(\x - \xground)))^2});
        \node[anchor=east, \groundmorsecolor] at(\xmax,\yground+1.6)  {\shortstack{Ground\\state}};
    
        \begin{scope}
            \clip plot[domain=\xground-2:\xmax,smooth,samples=50,variable=\x] ({\x},{\yground + \De*(1 - exp(-(\x - \xground)))^2}) -- (\xmax,\ymax) -- (0,\ymax) -- cycle;
            \foreach  \n in {0,1,2,3,4,5,6,7,8,9,10} {
                \pgfmathsetmacro\y{\yground + (2*\lambda - \n - 0.5)*(\n + 0.5)/\lambda^2*\De}  
                \draw[\groundmorsecolor] (0,\y) -- (\xmax,\y);
            }
        \end{scope}
    
        \foreach  \n/\label in {0/$v''\!=0$, 1/$v''\!=1$, 2/$v''\!=2$, 3/$v''\!=3$} {
            \pgfmathsetmacro\y{\yground + (2*\lambda - \n - 0.5)*(\n + 0.5)/\lambda^2*\De}  
            \pgfmathsetmacro\xlabel{\xground - 1.1*ln(1 - sqrt((\y - \yground)/\De))}
            \node[anchor=west, \groundmorsecolor] at ({\xlabel + 0.06}, \y) {\footnotesize \label};
        }
    
        \begin{scope}
            \clip plot[domain=0:\xmax,smooth,samples=50,variable=\x] ({\x},{\yground+\De*(\x - \xground)^2}) -- (\xmax,\ymax) -- (0,\ymax) -- cycle;
            \foreach \n in {0,1,2,3,4,5,6,7} {
                \pgfmathsetmacro\y{\yground + (\n + 0.5)/2}  
                \draw[\groundharmcolor] (0,\y) -- (\xmax,\y);
            }    
        \end{scope}
    \end{scope}

    \draw[domain=\xexc-2:\xmax,smooth,variable=\x,\excmorsecolor,thick] plot ({\x},{\yexc+\De*(1-exp(-(\x-\xexc)))^2});
    \draw[domain=0:\xmax,smooth,variable=\x,\excharmcolor,thick] plot ({\x},{\yexc+\De*(\x-\xexc)^2});
    \node[anchor=east, \excmorsecolor] at (\xmax,\yexc+1.6)  {\shortstack{Excited\\state}}; 
    \node[anchor=east, \excmorsecolor] at (\xmax,\yexc+\De+0.1)  {\footnotesize Morse potential}; 
    \node[anchor=west, \excharmcolor] at (\xexc+1.2,\ymax-0.3)  {\footnotesize \shortstack{Harmonic\\potential}}; 
    
        \begin{scope}
            \clip plot[domain=\xexc-2:\xmax,smooth,samples=50,variable=\x] ({\x},{\yexc + \De*(1 - exp(-(\x - \xexc)))^2}) -- (\xmax,\ymax) -- (0,\ymax) -- cycle;
            \foreach \n in {0,1,2,3,4,5,6,7,8,9,10} {
                \pgfmathsetmacro\y{\yexc + (2*\lambda - \n - 0.5)*(\n + 0.5)/\lambda^2*\De}  
                \draw[\excmorsecolor] (0,\y) -- (\xmax,\y);
            }
        \end{scope}
        
        \foreach  \n/\label in {0/$v'\!=0$, 1/$v'\!=1$, 2/$v'\!=2$, 3/$v'\!=3$} {
            \pgfmathsetmacro\y{\yexc + (2*\lambda - \n - 0.5)*(\n + 0.5)/\lambda^2*\De}  
            \pgfmathsetmacro\xlabel{\xexc - 1.06*ln(1 - sqrt((\y - \yexc)/\De))}
            \node[anchor=west, \excmorsecolor] at ({\xlabel + 0.1}, \y) {\footnotesize \label};
        }
    
        \begin{scope}
            \clip plot[domain=0:\xmax,smooth,samples=50,variable=\x] ({\x},{\yexc+\De*(\x - \xexc)^2}) -- (\xmax,\ymax) -- (0,\ymax) -- cycle;
            \foreach \n in {0,1,2,3,4,5,6,7} {
                \pgfmathsetmacro\y{\yexc + (\n + 0.5)/2}  
                \draw[\excharmcolor] (0,\y) -- (\xmax,\y);
            }    
        \end{scope}

    
    \draw[Goldenrod, ->, decorate, decoration={snake, amplitude=0.3mm, segment length=3mm}, thick] (\xground+0.02, {\yground + (2*\lambda - 0.5)*0.5/\lambda^2*\De})
    -- (\xground+0.02, {\yexc + (2*\lambda - \ntx - 0.5)*(\ntx + 0.5)/\lambda^2*\De});
    \draw[->, decorate, decoration={snake, amplitude=0.3mm, segment length=3mm}, thick] (\xground, {\yground + (2*\lambda - 0.5)*0.5/\lambda^2*\De})
    -- (\xground, {\yexc + (2*\lambda - \ntx - 0.5)*(\ntx + 0.5)/\lambda^2*\De});
   \node[anchor=east, rotate=90, Goldenrod] at (\xground-0.24, \yexc-0.0) {Excitation};
    \node[anchor=east, rotate=90] at (\xground-0.25, \yexc-0.0) {Excitation};
    
    \draw[Goldenrod, ->, decorate, decoration={snake, amplitude=0.3mm, segment length=3mm}, thick] 
       (\xexc+0.02, {\yexc + (2*\lambda - 0.5)*0.5/\lambda^2*\De})
    -- (\xexc+0.02, {\yground + (2*\lambda - \ntm - 0.5)*(\ntm + 0.5)/\lambda^2*\De});
    \draw[->, decorate, decoration={snake, amplitude=0.3mm, segment length=3mm}, thick]
       (\xexc+0, {\yexc + (2*\lambda - 0.5)*0.5/\lambda^2*\De})
    -- (\xexc+0, {\yground + (2*\lambda - \ntm - 0.5)*(\ntm + 0.5)/\lambda^2*\De});
    \node[rotate=270, Goldenrod] at (\xexc+0.27, \yexc-0.72) {Emission};
    \node[rotate=270] at (\xexc+0.26, \yexc-0.72) {Emission};
    
    \draw[dotted, thick,  ->] (\xground, {\yexc + (2*\lambda - \ntx - 0.5)*(\ntx + 0.5)/\lambda^2*\De})
    -- (\xexc-0.01, {\yexc + (2*\lambda - 0.5)*0.5/\lambda^2*\De + 0.03});

\end{scope}

\pgfmathsetmacro\middle{(\yground + \yexc)/2 + (2*\lambda - 0.5)*0.5/\lambda^2*\De/2}
\def\nupos{0.3}
\draw[dotted, \groundmorsecolor, thick] 
(\xground-0.3, {\yground + (2*\lambda - 0.5)*0.5/\lambda^2*\De}) -- (0, {\yground + (2*\lambda - 0.5)*0.5/\lambda^2*\De});
\draw[dotted, \excmorsecolor, thick]
(\xexc-0.3, {\yexc + (2*\lambda - 0.5)*0.5/\lambda^2*\De}) -- (0, {\yexc + (2*\lambda - 0.5)*0.5/\lambda^2*\De});
\draw[->] (\nupos, \middle+0.2) -- (\nupos, {\yexc + (2*\lambda - 0.5)*0.5/\lambda^2*\De});
\draw[->] (\nupos, \middle-0.2) -- (\nupos, {\yground + (2*\lambda - 0.5)*0.5/\lambda^2*\De});
\node[] at (\nupos, \middle) {$\nu_{00}$};

\node[] at (-0.3, \yground) {$E_0$};
\draw[] (-0.12, \yground) -- (0, \yground);
\node[] at (-0.3, \yexc) {$E_e$};
\draw[] (-0.12, \yexc) -- (0, \yexc);

\draw[dotted, thick, \groundmorsecolor] (\xground, \yground) -- (\xground, 0);
\draw[dotted, thick, \excmorsecolor] (\xexc, {\yground + (2*\lambda - \ntm - 0.5)*(\ntm + 0.5)/\lambda^2*\De}) -- (\xexc, 0);
\node[] at (\xground/2+\xexc/2, 0.2) {$\Delta \mathbf{R}$};

\end{tikzpicture}}
    \caption{
    Franck–Condon diagram illustrating vertical excitation from $v'' = 0$ of the electronic ground state to $v' = 4$ of the excited state, followed by non-radiative vibrational relaxation to $v' = 0$ (dotted arrow), and subsequent emission to $v'' = 2$ of the ground state. The ground and excited state potential energy surfaces (PES) are shown as Morse potentials, overlaid with their harmonic approximations. Since the equilibrium geometries of ground and excited state differ by $\Delta\mathbf{R}$, an electronic excitation results in a geometry that does not resemble the relaxed shape of the excited state as much as certain vibrationally excited states. This makes transitions to those states more likely. The energy difference $\tilde{\nu}_{00}$ between the vibrational ground states $(v',v''=0)$ is called the zero-zero transition energy, and should not be confused with the difference $E_e-E_0$ between the electronic potential minima of the excited and ground states.
    }

    \label{fig:fc_diagram}
\end{figure}
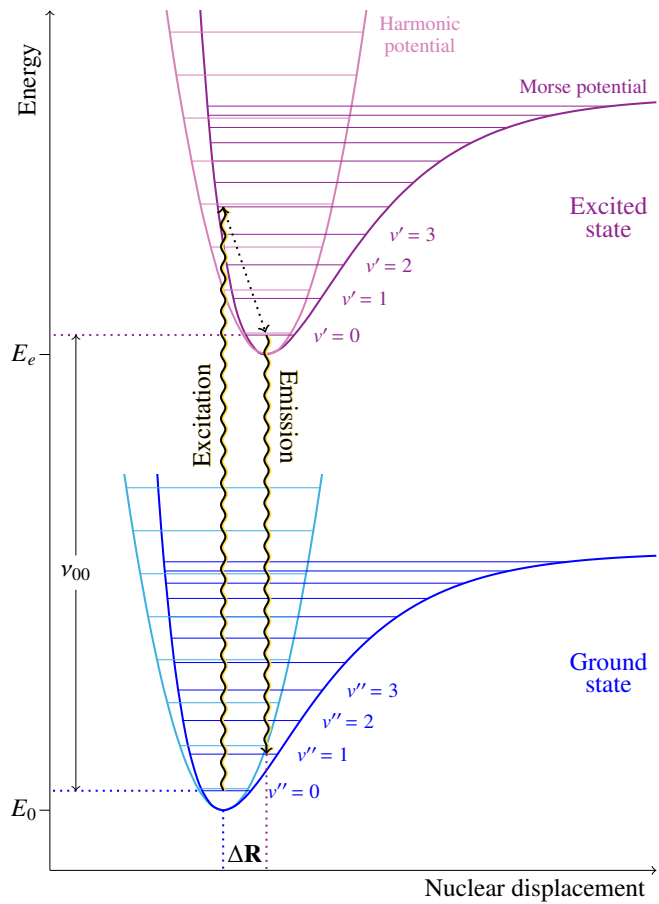

The foundation of most theoretical treatments of vibronic spectra is the Born–Oppenheimer approximation \cite{born1927}, which exploits the fact that electrons are much lighter, and thus move on a much faster timescale, than the nuclei.  Consequently, the total wavefunction of a molecule can be to a good approximation written as a product of an electronic and nuclear (vibrational) wavefunctions:
\[
\Psi(\mathbf{r}, \mathbf{R}) \approx \psi_{\text{el}}(\mathbf{r}; \mathbf{R}) \cdot \chi(\mathbf{R}),
\]
where \( \mathbf{r} \) and \( \mathbf{R} \) are electronic and nuclear coordinates, respectively. As a result, when a molecule transitions to a different electronic state, 
the nuclei remain effectively stationary and --- figuratively speaking --- suddenly find themselves on a new potential energy surface (PES), which describes the potential energy as a function of nuclear geometry in the new electronic state.
On a potential energy diagram such as Fig.~\ref{fig:fc_diagram}, such transitions are represented by vertical arrows; they are therefore known as \emph{vertical transitions}.

This observation forms the basis of the \textit{Franck--Condon principle} \cite{franck_elementary_1926, condon_theory_1926}, which states that the probability of a vibronic transition can be to a good approximation determined from the overlap between the vibrational wavefunctions of the initial and final electronic states. The larger this overlap, the more intense the corresponding spectral line. Formally, the transition intensity is obtained by computing the transition dipole moment $\mu_{if}$ between two vibronic states
\[
\mu_{if} = \iint \chi_f^*(\mathbf{R}) \psi_{\text{el}}^f(\mathbf{r}; \mathbf{R}) \, \hat{\mu}(\mathbf{r}, \mathbf{R}) \, \psi_{\text{el}}^i(\mathbf{r}; \mathbf{R}) \chi_i(\mathbf{R}) \, d\mathbf{r}\, d\mathbf{R},
\]
where \( \psi_{\text{el}}^{i,f} \) and \( \chi_{i,f} \) are the electronic and vibrational wavefunctions of the initial and final states, respectively, and \( \hat{\mu} \) is the electric dipole operator.

Under the \textit{Franck--Condon approximation}, it is assumed that the electronic transition dipole moment does not vary significantly with nuclear coordinates, i.e., \( \hat{\mu}(\mathbf{r}, \mathbf{R}) \approx \hat{\mu}(\mathbf{r}) \). This allows the integral to be factorized into electronic and vibrational parts:
\[
\mu_{if} \approx \left( \int \psi_{\text{el}}^f(\mathbf{r})\, \hat{\mu}(\mathbf{r})\, \psi_{\text{el}}^i(\mathbf{r})\, d\mathbf{r} \right)
\cdot
\left( \int \chi_f^*(\mathbf{R}) \chi_i(\mathbf{R})\, d\mathbf{R} \right).
\]

The first factor corresponds to the purely electronic transition moment (constant across the spectrum), while the second — the vibrational overlap usually referred to as the Franck-Condon factor — governs the relative intensities of individual vibronic transitions. The transition intensity $I_{if}$ is then given by
\[
I_{if} \propto |\mu_{if}|^2 \propto \left| \int \chi_f^*(\mathbf{R}) \chi_i(\mathbf{R})\, d\mathbf{R} \right|^2.
\]
This approximation works well for electronically allowed transitions in which the transition dipole is relatively insensitive to small displacements of the nuclei.

\subsubsection{Potential Energy Surfaces and the Harmonic Model}

To compute the vibrational wavefunctions needed to evaluate the Franck--Condon integrals, one must solve the nuclear Schrödinger equation on the PES associated with each electronic state. These PESs describe how the electronic energy varies as a function of nuclear geometry (as shown schematically in Fig.~\ref{fig:fc_diagram}) and are typically approximated near the equilibrium by a low-order Taylor expansion in the nuclear displacements.

Retaining only the quadratic terms in this expansion yields the \emph{harmonic approximation} \cite{wilson_molecular_1980}, in which the PES takes the form of a multidimensional paraboloid (shown as lighter-colored parabolas in Fig.~\ref{fig:fc_diagram}). In this model, the nuclear degrees of freedom decouple into independent harmonic oscillators, each corresponding to a \emph{normal mode} of vibration. The vibrational energy levels and wavefunctions can then be obtained analytically, and the associated overlap integrals evaluated in closed form.

In practice, normal modes are obtained by diagonalizing the mass-weighted Hessian matrix of second derivatives of the PES at the equilibrium geometry with respect to nuclear displacements~\cite{banwell1994, g16}.
Each eigenvector \( \mathbf{D}_m \), known as a displacement vector, defines a normal mode \( m \) of vibration: a coherent motion in which all atoms oscillate sinusoidally, with \( \mathbf{D}_m \) specifying the direction and relative amplitude of each atom’s displacement.
The corresponding eigenvalue is used to compute the vibrational frequency \( \nu_m \), which determines the energy spacing between quantized vibrational levels of the associated harmonic oscillator and is typically reported in units of cm$^{-1}$ as a wavenumber $\tilde{\nu}=\nu/c$, where $c$ is the speed of light.

Because of its simplicity and computational efficiency, the harmonic approximation forms the foundation of most practical vibronic simulations.
Alternative time-dependent approaches compute spectral intensities from autocorrelation functions~\cite{heller_timedependent_1975, mukamel_principles_1995}, bypassing the need for explicit PES construction. Statistical Franck–Condon models can be employed to reproduce band shapes without resolving individual transitions~\cite{Li_2017}, which is useful for large systems where fine vibronic structure is not resolved experimentally.

\subsubsection{Herzberg--Teller Approximation}

The Franck--Condon approximation assumes that the electronic transition dipole moment does not depend on the nuclear geometry. While this holds well for many allowed transitions, it breaks down in cases where the transition dipole varies significantly with nuclear displacement—most notably in electronically forbidden transitions, where the dipole moment is zero at the equilibrium geometry.

To account for such cases, the dipole operator can be expanded in a Taylor series about the equilibrium geometry \( \mathbf{R}_0 \) of the final state,
\[
\hat{\mu}(\mathbf{r}, \mathbf{R}) \approx \hat{\mu}(\mathbf{r}, \mathbf{R}_0) + \sum_m \left( \frac{\partial \hat{\mu}}{\partial Q_m} \right)_{\mathbf{R}_0} Q_m + \cdots.
\]
Here, the \( Q_m \) are mass-weighted normal coordinates in the basis spanned by the displacement vectors \(\mathbf{D}_m\).

The first term recovers the Franck--Condon approximation, while the linear term gives rise to additional contributions to the transition dipole — known as \emph{Herzberg--Teller coupling} \cite{herzberg_molecular_1950, herzberg_schwingungsstruktur_1933}. Physically, this term captures the fact that, even if a transition between equilibrium geometries is symmetry-forbidden, small nuclear displacements along a vibrational mode can transiently lower the symmetry of the molecule, making the transition weakly allowed. Herzberg--Teller coupling thus provides a mechanism allowing vibrational motion to give intensity to  otherwise forbidden electronic transitions.
This effect is particularly important in high-symmetry systems (e.g., benzene), where the lowest-energy electronic transitions are dipole-forbidden but become weakly allowed through Herzberg--Teller coupling. 

All computed spectra presented later in this paper include both Franck–Condon and Herzberg–Teller contributions, as evaluated by Gaussian's \texttt{fcht} implementation.

\subsubsection{Duschinsky Rotation}

In the harmonic approximation, each electronic state has its own set of normal modes, defined at its respective equilibrium geometry. Since the geometries and PES generally differ between electronic states, the basis sets of normal mode vectors \( \{ \mathbf{D}^{(i)}_m \} \) and \( \{ \mathbf{D}^{(f)}_m \} \) of the initial and final electronic states are typically not aligned.

This discrepancy must be accounted for when evaluating Franck–Condon and Herzberg–Teller integrals, which requires expressing vibrational wavefunctions in a common coordinate system. The associated change of basis — known as the \emph{Duschinsky transformation} \cite{duschinsky_zur_1937} — expresses the mass-weighted normal coordinates of the final state \( \mathbf{Q}^{(f)} \) in terms of those of the initial state \( \mathbf{Q}^{(i)} \),
\[
\mathbf{Q}^{(f)} = \mathbf{J} \mathbf{Q}^{(i)} + \mathbf{K},
\]
where \( \mathbf{J} \) is the Duschinsky rotation matrix that mixes modes, and \( \mathbf{K} \) is the displacement vector between the equilibrium geometries of the two electronic states, both expressed in the initial-state normal mode basis.

Neglecting the Duschinsky rotation can introduce systematic errors in the vibrational overlap integrals, especially when the geometry change between states is significant.
Modern vibronic simulation programs, including Gaussian's \texttt{fcht} implementation, account for the Duschinsky transformation when evaluating Franck--Condon and Herzberg--Teller integrals \cite{santoro_effective_2007, barone_fully_2009, g16}.

\subsubsection{Anharmonicity}

The harmonic approximation provides an efficient framework for modeling vibrational motion, but it only remains accurate near the equilibrium geometry, where PES tend to be well-described by a quadratic potential. In reality, molecular vibrations are anharmonic: the restoring force weakens at large bond stretches, leading to a Morse-like potential \cite{morse_diatomic_1929}; vibrational energy levels become unequally spaced, and coupling between modes can occur when vibrational motion deviates from the harmonic normal-mode directions \cite{herzberg_molecular_1950, wilson_molecular_1980}. These effects --- illustrated in Fig.~\ref{fig:fc_diagram} --- become particularly important at higher vibrational excitations and in modes involving light atoms, such as X--H stretches, and even more so in hydrogen-bonded clusters \cite{temelso2011}.

Accurate treatment of these effects requires detailed modeling of the multidimensional PES. For very small molecules, this is feasible: Morse-like potentials describe diatomics \cite{leroy_2009, qadeer_morselike_2022}, while multidimensional analytic PES fitted to ab initio points can be constructed for up to ~15 internal degrees of freedom \cite{quack_PES_1991, vendrell_15dI_2007}, and solved in terms of full-dimensional wavefunctions \cite{zhang_exact_1995, vendrell_15dII_2007}.
To account for anharmonicity in medium-sized molecules, high-dimensional approximate PES are used in approaches such as the Vibrational Self-Consistent Field (VSCF) \cite{vscf_bowman_1978, vscf_chaban_1999}, Vibrational Configuration Interaction (VCI) \cite{bowman_variational_2008, christiansen_selected_2012}, Multi-Configuration Time-Dependent Hartree (MCTDH) \cite{meyer_multi-configurational_1990, bowman_multimode_2003, yu_fulldim_2015}, and self-consistent-charge DFTB molecular dynamics \cite{nishimura_2014} methods.
Local-mode approaches \cite{cheng_localmode_2014} further reduce the complexity by isolating strongly anharmonic or localized vibrations (e.g., X–H stretches) and treating them separately, enabling application to large systems such as peptides \cite{panek_localmode_2016}.
The need to construct a global PES can be avoided entirely by on-the-fly semiclassical methods such as SC-IVR and MC-SCIVR, which produce vibrational spectra directly from ab initio trajectories \cite{ceotto_multiple_2009, conte_reproducing_2013, aieta_anharmonic_2020}.
While accurate, these methods are computationally demanding and rarely applied in large-scale vibronic analyses.

For moderately anharmonic but near-rigid systems, vibrational perturbation theory (VPT2) \cite{barone_anharmonic_2005} offers a computationally efficient alternative based on a quartic Taylor expansion of the PES around equilibrium geometry.
In Gaussian \cite{g16}, this is enabled via the \texttt{freq=anharm} keyword, which provides anharmonically corrected vibrational wavenumbers and zero-point energies. However, as currently implemented, VPT2 is restricted to the electronic ground state and cannot be combined with excited-state vibronic simulations such as those using the \texttt{fcht} keyword in Gaussian.

In practice, the use of computationally expensive anharmonic methods can be often disregarded, as anharmonic effects are largely systematic \cite{irikura_scaling_2009}, and can be effectively accounted for by applying empirical scaling factors \cite{jacobsen_notWorthwhile_2013}. This idea was introduced by Pople et al. \cite{pople_1981}, who observed that an effective scaling factor of 0.8929 brings the calculated harmonic wavenumbers at the HF/3-21G level of theory close to the experimental values.
Scaling factors can either be applied to the force constants, or directly to the vibrational wavenumbers, using a variety of strategies \cite{palafox_2018}. Rather than using a single global scaling factor, improved accuracy can be achieved by applying distinct scaling factors to ZPE and fundamental wavenumbers  \cite{irikura_scaling_2009, grev_zpeVSfundamentalScaling_1991, barone_beyond_2004}, or even to low- and high-wavenumber vibrational modes \cite{scott_leastsquareScaling_1996, bauschlicher_PAH_1996, wheeler_2004, johnson_2010}.
Scaling factors typically fall within the range 0.9--1.0, and have been reported for a wide variety of method and basis set combinations \cite{scott_leastsquareScaling_1996, wheeler_2004, witek2004, merrik_factors_2007, alecu_2010, nist_cccbdb_scaling}.

High-wavenumber scaling factors tend to be smaller, as X–H stretching modes dominate this region and follow broad, strongly anharmonic Morse-like potentials. Meanwhile, other molecular deformations involving heavier atoms typically remain closer to harmonic behavior. To apply separate scaling factors in practice, a cut-off wavenumber has been introduced in the literature to distinguish low- and high-wavenumber modes. This threshold is somewhat arbitrary and molecule-dependent --- choices range from $1000~\text{cm}^{-1}$ \cite{wheeler_2004}, to $2700~\text{cm}^{-1}$ \cite{johnson_2010}. An alternative is the wavenumber-linear scaling (WLS) approach \cite{yoshida_wls_2002}, which linearly decreases the scaling factor with the wavenumber, as ${\tilde{\nu}_\text{corr}}/{\tilde{\nu}_\text{calc}} = 1-\lambda \tilde{\nu}_\text{calc}$.

While effective in many cases, this heuristic may misclassify modes near the cut-off wavenumber, particularly in complicated systems, where vibrations of different types (such as X-H stretches, out-of-plane bends, and other deformations) can appear in overlapping wavenumber ranges. It would be more accurate to determine the type of each vibrational mode separately, but manual classification is feasible only for small molecules (e.g. \cite{bauschlicher_PAH_1996}), becoming quickly impractical for systems with many vibrational degrees of freedom. 

For this reason, \spec implements an automatic categorization of each vibrational mode based on its displacement vector. This classification enables the program to apply type-specific scaling factors when computing corrected wavenumbers for individual transitions. The resulting correction scheme is detailed in Subsection~\ref{subsec:specdisplay}.

\subsubsection{Peak Shape and Broadening}

Unlike the idealized stick theoretical spectra, experimental spectra exhibit broadened peaks. This broadening arises from several physical and instrumental effects: finite instrumental resolution, thermal population distributions, finite lifetimes of excited states (\textit{via} the time–energy uncertainty principle), and collisional broadening in the gas or condensed phase \cite{atkins2002}. As a result, theoretical stick spectra composed of discrete transitions must be convoluted with lineshape functions to resemble experimental lineshapes.

Two common models for spectral lineshapes are the Gaussian and Lorentzian functions. Gaussian profiles typically arise from inhomogeneous or Doppler broadening, while Lorentzian shapes reflect homogeneous broadening mechanisms such as finite excited-state lifetimes \cite{herzberg_molecular_1950}. In practice, Lorentzian profiles are widely used in vibronic simulations due to their narrower central regions and longer tails, which often provide better match with experimental spectra, particularly in condensed-phase or low-temperature measurements \cite{barone_fully_2009}.

The Lorentzian lineshape function is defined as
\begin{equation}\label{eq:lorentzian}
L_w(\tilde{\nu}) = \frac{1}{1 + \left( \frac{\tilde{\nu}}{w} \right)^2},
\end{equation}
where \( w \) is the half-width at half-maximum (HWHM). The function is scaled such that its maximum is $L_w(0)=1$.

The full simulated spectrum is computed by convoluting each discrete transition with the lineshape function:
\begin{equation}\label{eq:specsum}
S(\tilde{\nu}) = \sum_t i_t \cdot L_w(\tilde{\nu} - \tilde{\nu}_{t}),
\end{equation}
where \( i_t \) is the intensity of transition \( t \), and \( \tilde{\nu}_t \) is the central wavenumber of the transition.

In most practical simulations, the value of \( w \) is chosen empirically to reflect the resolution and broadening observed in the experimental spectrum \cite{barone_fully_2009}. In \spec, this parameter is user-adjustable in the GUI as one of the tools for aligning the shape of simulated spectra with experimental observations.

\section{Software Workflow and Functionality} \label{sec:workflow}

\spec is a standalone desktop application written in Python, distributed both as source code (available at \href{https://github.com/giogina/SpectraMatcher/}{github.com/giogina/SpectraMatcher/}) and as compiled binaries for Windows and Linux. These binaries can be found in the \href{https://github.com/giogina/SpectraMatcher/tree/main/windows\_installer}{windows\_installer/} and \href{https://github.com/giogina/SpectraMatcher/tree/main/linux\_installer}{linux\_installer/} directories of the repository.

A detailed user manual and documentation is available at \href{https://spectramatcher.gitbook.io/}{spectramatcher.gitbook.io/}. The Windows executable supports Windows~7 and later. The Linux executable requires glibc version~2.31 or newer, meaning that it is compatible with distributions such as Ubuntu~20.04+, Debian~11+, and Fedora~32+. Users on other platforms can run the program from source, provided that Python~3.7 or later and the required dependencies are installed.

\subsection{Overview of Workflow}

\spec guides users through a structured workflow consisting of four main stages: \textit{data import}, \textit{spectral manipulation}, \textit{peak matching}, and \textit{output generation}. Each stage provides interactive parameter control and immediate visual feedback, enabling efficient and reproducible comparison between experimental and computed vibronic spectra.

Upon starting a new project, the user imports a set of experimental spectra alongside frequency and vibronic output files from \textit{ab initio} calculations. These files are automatically parsed and checked for consistency in molecular identity, level of theory, and geometry. \spec then constructs computed spectra by convoluting Franck--Condon / Herzberg--Teller transitions with a Lorentzian line shape and applying type-specific anharmonic correction factors. In parallel, peaks are automatically detected in the experimental spectra using configurable prominence and width thresholds.

The resulting spectra can be interactively adjusted by modifying parameters such as zero-zero transition energy, intensity scaling, and peak half-width. This allows the user to visually align computed spectra with experimental observations. The software then automatically assigns computed to experimental peaks using a customizable algorithm based on wavenumber proximity and relative intensity thresholds.

Finally, the table of peak assignments and adjusted spectra can be exported in a variety of formats suitable for further analysis or inclusion in publications. This workflow reduces the manual effort required for vibronic spectral analysis and provides control over key modeling parameters.

\subsection{Project Setup and File Import}

When a new \spec project is started, the corresponding project file (\texttt{.spm}) is created, into which progress --- including imported files, spectrum manipulations, and match settings --- are stored. This allows users to revisit their work without losing progress. A robust auto-save mechanism protects users from accidental data loss.

Files are imported \textit{via} an integrated file manager, which provides a preview interface summarizing file types and relevant metadata --- such as (in the case of Gaussian output files) job type, completion status, molecular formula, method, basis set, and zero-zero transition energy. Outputs are flagged if they contain negative frequencies, are incomplete, or the computation terminated abnormally.

Gaussian \texttt{.log} files are further parsed if they contain either frequency calculations or vibronic simulations. From frequency calculation outputs, for each normal mode $m$, the program extracts the vibrational wavenumbers $\tilde{\nu}_m$ (in cm$^{-1}$), the displacement vectors $\mathbf{D}_m$ and the symmetry classification. If the computation was performed with the \texttt{freq=hpmodes} keyword, high-precision displacement vectors (five decimal places) are used; otherwise, low-precision data (two decimal places) is extracted.

From vibronic calculation outputs, \spec parses, for each transition $t$, the transition wavenumber $\tilde{\nu}_t$, intensity $i_t$, and the set of involved vibrational modes $m$, along with their corresponding vibrational quantum numbers $n_{t,m}$. These quantities are used to construct stick spectra representing Franck--Condon and Herzberg--Teller transitions.

In addition, from all Gaussian files, the program extracts the Cartesian molecular geometry in the form of coordinate vectors $\mathbf{R}_x$, $\mathbf{R}_y$, $\mathbf{R}_z$ and a vector of atomic numbers $\mathbf{A}$, as well as the total electronic and zero-point energies.

Experimental spectra can be provided in plain-text or spreadsheet formats. Columns for wavenumbers and intensities are automatically recognized based on headers or file structure, but users may manually reassign them after import, if needed.

From the files listed in the file manager, users may either select files manually, or use the auto-import function, which detects and imports the largest compatible set of files with matching molecular identity, electronic state, and computational method. In both modes, \spec performs internal consistency checks to prevent mismatches in molecular geometry or zero-point energy.

After parsing the necessary input files, \spec classifies vibrational modes and prepares the data for spectral analysis.

\subsection{Analysis of Experimental Spectra}
In all provided experimental spectra, peak detection is performed automatically based on user-defined thresholds for minimum prominence and peak width. To ensure robust detection in noisy spectra, this process operates on a smoothed version of the original data, using a moving average filter. Users may also add or remove peaks manually through the graphical interface. The resulting experimental peak list forms the basis for spectral matching and assignment, as described in Subsection~\ref{subsec:matching}.

\subsection{Mode Classification for Wavenumber Scaling}
\label{subsec:modeclassification}

To enable anharmonic corrections through vibration-type-specific wavenumber scaling, each vibrational mode is automatically classified into one of three categories: out-of-plane (OOP) bends, X--H stretches, and other deformations.

A molecule is considered planar if there is one Cartesian axis \( o \in \{x, y, z\} \) along which the maximum atomic displacement is less than 1~\AA.
This heuristic is sufficient in practice, as planar molecules are aligned with a Cartesian plane during Gaussian computations.
For planar molecules, a vibrational mode \( m \) is classified as an OOP bend if its (normalized) displacement vector \( \mathbf{D}_m \) has significant amplitude in the out-of-plane direction $o$:
\[
\| \mathbf{D}_{mo} \|^2 > 0.9,
\]
where \( \mathbf{D}_{mo} \) denotes the projection of \( \mathbf{D}_m \) onto the axis  \( o \).

Hydrogen OOP bending is sometimes considered a separate subclass \cite{hony2001ch}. However, in test cases involving pyrene, naphthalene, and ovalene, all OOP bends were found to include hydrogen displacement to varying degrees. Since these modes carry negligible intensity in the examined vibronic spectra, further subclassification was deemed unnecessary.

A vibrational mode is classified as an X--H stretch if it involves stretching motion along a bond between a hydrogen atom $h$, with position $\mathbf{r}_h$ and displacement $\mathbf{d}_h$, and a bonded heavy atom $a$, with position $\mathbf{r}_a$ and displacement $\mathbf{d}_a$, such that
\[
(\mathbf{r}_a - \mathbf{r}_h) \cdot (\mathbf{d}_a - \mathbf{d}_h) > 0.2~\text{\AA}.
\]

All vibrational modes that satisfy neither of the above criteria are assigned to the category of ``other deformations''. These include stretches between heavy atoms and in-plane bends of hydrogen and/or heavy atoms.

\subsection{Gaussian and Mulliken Vibrational Mode Labels}
\label{subsec:modelabels}

Vibrational modes can be labeled in the plot either by their index in the Gaussian output or using \textit{Mulliken} symmetry notation, which is commonly employed in spectroscopic literature.

\spec assigns Mulliken labels to vibrational modes sequentially, beginning with 1, primarily in order of symmetry class, as described below, and then, within each symmetry class, in order of descending wavenumber.

The symmetry class of each mode is taken directly from the Gaussian \texttt{freq} output. By default, symmetry classes are arranged in an order consistent with typical spectroscopic convention. For example, in \texttt{D\textsubscript{2h}} symmetry, the default order is
\[
\texttt{AG},\ \texttt{B1G},\ \texttt{B2G},\ \texttt{B3G},\ \texttt{AU},\ \texttt{B1U},\ \texttt{B2U},\ \texttt{B3U}.
\]
Placeholder symmetry classes (e.g.,~\texttt{?A}, etc.) are appended at the end of this list in alphabetical order. This default ordering can be manually adjusted in the user interface if needed.

\subsection{Display and Manipulation of Spectra}\label{subsec:specdisplay}

Once all required input files are loaded, SpectraMatcher generates overlaid emission and excitation spectra for both the experimental data and computed transitions.

The experimental spectrum is plotted as-is. Peaks are detected automatically based on adjustable prominence and width thresholds, which the user can tune within the interface. Alternatively, peak markers can be added or removed manually.

The theoretical spectrum for each excited state is obtained from the outputs of the Gaussian calculations as follows.
For each transition $t$, let us denote by $T = \{m_1, m_2, ...\}$ the set of involved vibrational modes $m_j$, and by $v_j$ their respective vibrational quantum numbers in the transition's final electronic state. Then, the corrected wavenumber of the transition is computed as
\begin{equation}\label{anharm_correction}
\tilde{\nu}_t = \sum_{m \in T} f_{\tau(m)} \, v_{m} \, \tilde{\nu}_{m},
\end{equation}
where $\tilde{\nu}_m$ denotes the harmonic wavenumber of mode $m$ (as obtained from the frequency calculation), and $f_{\tau(m)}$ is the user-defined scaling factor associated with the vibrational type $\tau(m)$ of $m$ (i.e., OOP bend, X--H stretch, or other; as obtained according to Subsection~\ref{subsec:modeclassification}).

The transitions $t$, with their corrected positions $\tilde{\nu}_t$ (Eq.~\eqref{anharm_correction}) and their intensities $i_t$ obtained from the Franck--Condon/Herzberg--Teller calculations, are then convoluted with a Lorentzian line shape (Eq.~\eqref{eq:lorentzian}) according to Eq.~\eqref{eq:specsum}.
The half-width \( w \) is set separately for excitation and emission spectra, and is initialized based on the half-width observed in the experimental spectrum, but remains user-adjustable.

By default, computed spectra are normalized to unit maximum intensity and shifted so that the zero-zero transition energy \( \tilde{\nu}_{00} \) aligns with the origin of the experimental spectrum’s relative wavenumber axis. Scaling factors and wavenumber shifts can be adjusted interactively; and individual spectra can be toggled on or off. Transitions in the computed spectrum may be annotated using Gaussian mode indices or Mulliken labels (compare Subsection~\ref{subsec:modelabels}). For each vibrational mode, an animation of the associated nuclear motion is available aid interpretation.

\subsection{Peak Matching}\label{subsec:matching}

To assist in the interpretation of complex spectra, \spec includes a novel automated algorithm for assigning corresponding peaks between the aligned experimental and computed spectra. The algorithm matches peaks in order of decreasing intensity. This anchors the spectra at their most prominent features before smaller, less significant peaks are considered, which improves robustness when dealing with complex or overlapping spectral features.

A pair of user-configurable thresholds is used to determine whether an experimental and computed peak can be matched:

\begin{itemize}
    \item The \textbf{wavenumber distance threshold} \( \tau_{\Delta \tilde{\nu}} \), which defines the maximum allowed difference in wavenumber \( \Delta \tilde{\nu} \) between matched peaks.
    \item The \textbf{relative intensity threshold} \( \tau_I \), which sets the minimum acceptable ratio between the intensities of the experimental and computed peaks.
\end{itemize}

The default value for \( \tau_{\Delta \tilde{\nu}} \) is 30~cm\(^{-1}\), while the default for \( \tau_I \) is 0.03. The latter prevents high-intensity experimental peaks from being matched to computed peaks with less than 3\% of their intensity. Setting \( \tau_I = 0 \) disables intensity filtering entirely, while \( \tau_I = 1 \) only allows matches between peaks of equal intensity.

If multiple experimental peaks satisfy both threshold criteria for a given computed peak, the best-matching one is selected. To this end, in \spec, we introduce the scoring function
\[
\frac{I_{\text{exp}}}{\Delta \tilde{\nu}^2}
\]
which favors peaks that are both intense and close in wavenumber.
Matched peaks are visualized using vertical lines in the overlaid spectra, and the resulting assignment table lists wavenumbers, intensities, and vibronic transition information for each matched pair.
For further algorithmic details, see Section~\ref{subsec:matching-algo}.

\subsection{Output and Export Options} \label{subsec:output}

\spec provides export functionality for both tabular and numerical data, facilitating further analysis and integration into scientific publications.

Matched peak assignments --- including wavenumber, intensity, and transition information --- can be exported as a plain-text table for import into external tools,  or as formatted \texttt{.docx} and \LaTeX{} tables for direct inclusion in reports or manuscripts.

Computed spectra can also be exported numerically as tab-separated files containing wavenumber and intensity values, enabling detailed post-processing or comparison with external datasets.

\section{Architecture and Implementation} \label{sec:implementation}

\texttt{SpectraMatcher} is implemented as a modular, desktop-based application with an emphasis on clarity, extensibility, and efficient data handling. The full source code is publicly available under the MIT license at:

\begin{center}
\texttt{https://github.com/giogina/SpectraMatcher}
\end{center}

The program is written in Python \cite{van2009python} (compatible with Python~3.7 and later) and structured according to the Model--View--ViewModel (MVVM) pattern, which separates interface logic from data models and enables a clean, maintainable architecture.
For distribution on Windows and Linux, the program is compiled into standalone executables using Nuitka \cite{hayen_nuitka_2025}, which converts the Python source code into optimized C++ and bundles all dependencies to enable execution without a Python installation.
The remainder of this section outlines the internal design patterns, data management strategies, and implementation of core features including file parsing, spectrum generation, and peak assignment.

\subsection{Program Architecture and Design Patterns}

The graphical interface is implemented using the \texttt{DearPyGui} library \cite{dearpygui}, selected for its efficient rendering and strong support for interactive, high-performance plotting. Interactive visualization is central to SpectraMatcher’s workflow, and \texttt{DearPyGui} provides real-time feedback during spectral manipulation and matching tasks.

Application state is managed by a central \texttt{Project} class, which holds references to all imported data, analysis parameters, and derived results. Projects are saved to disk as standalone files and can be reopened or updated at any time. To avoid data corruption during concurrent operations, file access is coordinated using a \texttt{PathLockManager} class that ensures safe read/write operations through file-level locking.

To maintain a responsive interface, long-running tasks such as Gaussian log parsing and spectrum plotting are executed asynchronously. The \texttt{AsyncManager} class coordinates these operations using Python’s \texttt{asyncio} and \texttt{threading} libraries. Tasks submitted through \texttt{AsyncManager.submit\_task()} can register observers that are notified upon completion, enabling reactive updates in the user interface or downstream components.

\spec adopts an event-driven architecture combined with the observer pattern. Internal events --- such as file imports, data updates, or spectrum modifications --- trigger registered callbacks across the system. This ensures that GUI elements and data representations remain synchronized in real time without requiring explicit polling or manual refreshes.

\subsection{Data Parsing and Validation}

The \texttt{DataFileManager} class manages the lifecycle of files and directories within a project. When a user adds files --- either via the system’s file dialog or through drag-and-drop --- the \texttt{DataFileManager} creates a new \texttt{File} object for each entry. Each object determines its type by inspecting the file extension or, if necessary, the file contents, and then initiates asynchronous parsing.

Experimental spectrum files are recognized by their format (\texttt{.csv}, \texttt{.tsv}, \texttt{.txt}, \texttt{.xlsx}, \texttt{.xls}, \texttt{.xlsm}, or \texttt{.ods}). These are parsed by the \texttt{ExperimentParser}, which extracts numeric arrays representing absolute and relative wavenumbers and corresponding intensities. Files with missing, malformed, or otherwise unparseable content are skipped.

For Gaussian log files (\texttt{.log}), the \texttt{GaussianParser} is used. In an initial pass, the parser identifies the type(s) of Gaussian job and checks for successful termination. Incomplete or currently running calculations --- due to errors, manual interruption, or lack of an “end of file” marker --- are omitted and marked with a red error icon in the file explorer. Files that contain completed frequency or vibronic analyses are marked with a green checkmark and parsed further.

From frequency and vibronic jobs, the parser extracts the molecular geometry and the sum of electronic and zero-point energy. For frequency calculations, vibrational wavenumbers, displacement vectors, and symmetry classifications are stored in instances of the \texttt{VibrationalMode} class. Each mode is then classified as an out-of-plane (OOP) bend, X--H stretch, or other deformation, as described in Section~\ref{subsec:modeclassification}.

To ensure consistency, Gaussian output files are grouped automatically if they correspond to the same excited state of the same molecule, based on geometry and energy matching. When importing manually, files that do not match the molecule, ground-state energy, or 0--0 transition energy of those already present in the project are rejected, and a corresponding error message is shown in the interface. This consistency check can be disabled.

For vibronic calculation files, the Franck--Condon/Herzberg--Teller transitions are extracted. Each transition’s wavenumber, intensity, and contributing vibrational modes are stored in an \texttt{FCPeak} object.

Due to the modular design of the parsing subsystem, support for additional quantum chemistry software packages can be added by implementing a parser class and registering it under a new \texttt{File.file\_type} value.

\subsection{Spectrum Plotting and Manipulation}

The plotting system in \spec is designed for high efficiency to ensure smooth, real-time feedback during spectrum manipulation. This is achieved through optimized \texttt{NumPy}-based array operations \cite{numpy} for constructing convoluted theoretical spectra, and by limiting re-computation to only those cases where essential parameters --- e.g. peak half-width or anharmonic correction factors --- have changed.

The interactive visualization of computed spectra is managed by the \texttt{SpecPlotter} class. Each instance is initialized with a specific set of parameters that require a full convolution: the peak half-width \( w \), the wavenumber step size \( \delta \tilde{\nu} \), the plotting range \( \tilde{\nu}_s \ldots \tilde{\nu}_e \), and the anharmonic correction factors. To avoid redundant computations, previously constructed plotters are cached using these parameter sets as keys.

When a new set of correction factors is applied, the \texttt{WavenumberCorrector} computes the anharmonically corrected transition positions \( \tilde{\nu}_t^{\text{corr}} \) according to Eq.~\eqref{anharm_correction}. If the half-width \( w \) or step size \( \delta \tilde{\nu} \) has changed, a new set of Lorentzian shape templates (“stamps”) \( L^\mu_w(\tilde{\nu}) \) is generated, centered at \( \mu = 0, 0.1, 0.2, \ldots, \delta \tilde{\nu} \), and stored as \texttt{NumPy} arrays.

To compute the final convoluted spectrum \( S(\tilde{\nu}) \), the following steps are performed: for each transition, the index \( \iota \) of \( \tilde{\nu}_t^{\text{corr}} \) in the spectrum’s wavenumber grid is determined, and a precomputed Lorentzian stamp \( L^\mu_w(\tilde{\nu}) \) is selected such that
\[
|\tilde{\nu}_s + \iota \delta \tilde{\nu} + \mu - \tilde{\nu}_t^{\text{corr}}|
\]
is minimized. The selected array is then truncated and aligned so that its maximum coincides with \( \tilde{\nu}_t^{\text{corr}} \), scaled by the transition intensity, and summed into the spectrum. This procedure yields \( S(\tilde{\nu}) \) in agreement with Eq.~\eqref{eq:specsum}.

Each local maximum in the resulting spectrum is associated with the list of individual transitions that contribute to it. These are compiled along with the corresponding peak label for display and analysis.

When parameters that uniformly affect the entire spectrum, such as the intensity scaling factor or wavenumber shift, are modified, the entire spectrum \texttt{NumPy} array is scaled or translated accordingly, avoiding recomputation. Similarly, composite spectra for multiple excited states can be computed efficiently by summing the corresponding intensity arrays.

Spectra are displayed using \texttt{DearPyGui}’s interactive plotting features, which support draggable points and lines. This enables the user to shift and scale spectra directly within the plot, reposition labels, and interactively explore transitions.

\subsection{Experimental Spectrum Preprocessing}
Experimental spectra are represented by the \texttt{ExperimentalSpectrum} class, which manages both the raw intensity data and the list of detected peaks. To improve robustness against noise, peak detection is performed on a smoothed version of the data, using a seven-point moving average filter. This preprocessing is used solely for the purpose of locating local maxima; the original unsmoothed data is preserved for display and export. Peak positions are then identified using a lightweight in-house implementation that replicates the behavior of \texttt{scipy.signal.find\_peaks()} \cite{2020SciPy-NMeth}, with user-configurable thresholds for prominence and width.

\subsection{Matching Algorithm}\label{subsec:matching-algo}

The matching of experimental and computed peaks is handled by the \texttt{Matcher} class. Let \( E = \{(x_e, y_e)\} \) denote the set of experimental peaks, and \( C = \{(x_c, y_c)\} \) the set of computed peaks, where \( x \) is the wavenumber and \( y \) the intensity. Computed peaks are simply defined as local maxima of the spectrum \texttt{NumPy} array.

Peaks are processed in order of descending intensity, starting with the most intense computed peak. Prioritizing high-intensity features improves the robustness of the matching, especially for complex spectra with overlapping or weak peaks.

A computed peak \( (x_c, y_c) \in C \) is eligible to be matched with an experimental peak \( (x_e, y_e) \in E \) if the following two criteria are satisfied:
\begin{align}
|x_c - x_e| &< \tau_{\Delta \tilde{\nu}}, \\
\min\left( \frac{y_c}{y_e}, \frac{y_e}{y_c} \right) &> \tau_I,
\end{align}
where \( \tau_{\Delta \tilde{\nu}} \) is the wavenumber distance threshold (default: 30~cm\(^{-1}\)), and \( \tau_I \) is the relative intensity threshold (default: 0.03). These parameters can be adjusted by the user.

If multiple experimental peaks satisfy both criteria, the one with the highest score
\begin{equation}
\text{score} = \frac{y_e}{(x_c - x_e)^2}
\end{equation}
is selected as the best match. This scoring heuristic favors experimental peaks that are both close in wavenumber and relatively intense.

Once a match is made, the corresponding experimental peak is marked as used and excluded from further consideration. The algorithm then proceeds to the next most intense computed peak, continuing until no further eligible matches are found.

The output is a list of matched peak pairs, each consisting of the computed peak, the matched experimental peak, and the associated vibronic transition label. This information is displayed in the assignment table and can be exported as described in Section~\ref{subsec:output}.

\subsection{Performance}

\spec is designed to remain responsive and efficient, even when processing large datasets consisting of numerous Gaussian output files. This is achieved through extensive use of caching, asynchronous parsing, and selective recomputation of spectra.

To benchmark typical performance, a representative dataset of 40 Gaussian log files --- comprising 20 frequency calculations and 20 vibronic calculations, with an average file size of 2.5~MB --- was parsed. All data preparation steps, including vibrational mode classification, transition extraction, and initial spectrum convolution, completed in under 3~seconds on a standard desktop machine (Intel Core i7, 3.6~GHz, 16~GB RAM, SSD storage).

Plotting the convoluted spectra for each excited state using Lorentzian profiles, as well as rendering interactive overlays within the GUI, remained consistently responsive. Redraw times were below 200~ms for individual plots.

By recomputing convoluted spectra only when necessary—such as when the peak half-width or anharmonic correction factors are changed—and applying direct scaling or translation otherwise, the software supports real-time spectrum manipulation even for large datasets.

This performance enables interactive analysis of large polyatomic systems involving thousands of vibronic transitions, on standard personal computing hardware.

\section{Case Study} \label{sec:case_study}

\begin{table*}[bht]
\centering
\label{tab:ovalene_emission_assingments}
\caption{Auto-generated table of assignments of experimental and simulated peaks for the ovalene emission spectrum; truncated for brevity to omit the wavenumber range $1000\text{ cm}^{-1}<\tilde{\nu}<3000\text{ cm}^{-1}$. For each computed peak, the wavenumber $\tilde{\nu}$ and intensity of the convoluted peak is given, together with information for each contributing vibronic transition, including both the harmonic wavenumbers $\tilde{\nu}$ and corrected wavenumbers $\tilde{\nu}^\text{corr}$. The vibrational frequency scaling factors applied here are $f_\text{X-H}=0.977$ for X-H stretching modes (most notably adjusting the peak at $3092\text{ cm}^{-1}$ to fit the experiment)  and $f_\text{others}=0.988$. \\}
\renewcommand{\arraystretch}{1.1}
\begin{tabular}{ c c c c c c c c c c c }
\toprule
\multicolumn{2}{c}{Experiment} & \multicolumn{2}{c}{Computed} & & \multicolumn{6}{c}{Transition}\\
\cmidrule(lr){1-2} \cmidrule(lr){3-4} \cmidrule(lr){5-11}
$\tilde{\nu}$ & intensity & $\tilde{\nu}$ & intensity & state & & $\tilde{\nu}$ & $\tilde{\nu}^\text{corr}$ & intensity & sym & type \\
\midrule
-2.4 & 0.209 & 0.0 & 0.109 & 2nd & $0^{\smash{\scriptstyle 0}}_{\smash{\scriptstyle 0}}$ & 0.0 & 0.0 & 0.108 &  &  \\
315.0 & 0.604 & 316.0 & 0.467 & 2nd & $45^{\smash{\scriptstyle 0}}_{\smash{\scriptstyle 1}}$ & 319.6 & 315.7 & 0.422 & $B_{3g}$ &  \\
 &  &  &  &  & $23^{\smash{\scriptstyle 0}}_{\smash{\scriptstyle 1}}$ & 326.8 & 322.9 & 0.058 & $A_{g}$ &  \\
419.3 & 0.026 & 422.0 & 0.093 & 2nd & $22^{\smash{\scriptstyle 0}}_{\smash{\scriptstyle 1}}$ & 427.3 & 422.2 & 0.084 & $A_{g}$ &  \\
710.5 & 0.064 & 712.0 & 0.050 & 2nd & $41^{\smash{\scriptstyle 0}}_{\smash{\scriptstyle 1}}$ & 720.1 & 711.5 & 0.037 & $B_{3g}$ &  \\
765.1 & 0.133 & 766.0 & 0.113 & 2nd & $19^{\smash{\scriptstyle 0}}_{\smash{\scriptstyle 1}}$ & 775.0 & 765.7 & 0.089 & $A_{g}$ &  \\
883.6 & 1.000 & 882.0 & 1.000 & 2nd & $40^{\smash{\scriptstyle 0}}_{\smash{\scriptstyle 1}}$ & 892.4 & 881.7 & 0.962 & $B_{3g}$ &  \\
917.8 & 0.169 & 914.0 & 0.283 & 2nd & $18^{\smash{\scriptstyle 0}}_{\smash{\scriptstyle 1}}$ & 929.5 & 918.3 & 0.164 & $A_{g}$ &  \\
\multicolumn{11}{c}{\vdots} \\
3095.4 & 0.022 & 3095.0 & 0.144 & 2nd & $25^{\smash{\scriptstyle 0}}_{\smash{\scriptstyle 1}}$ & 3167.6 & 3094.7 & 0.067 & $B_{3g}$ & X-H \\
 &  &  &  &  & $3^{\smash{\scriptstyle 0}}_{\smash{\scriptstyle 1}}$ & 3168.3 & 3095.4 & 0.061 & $A_{g}$ & X-H \\
\bottomrule
\end{tabular}
\renewcommand{\arraystretch}{1.0}
\end{table*}

We highlight the key functionalities of \spec through the analysis of the dispersed fluorescence and fluorescence excitation spectra of ovalene (C$_{32}$H$_{14}$), a planar, \textit{peri}-condensed PAH. C$_{32}$H$_{14}$ has drawn attention in chemistry and related fields \textit{inter alia} as functional material for nanoelectronics and organic light-emitting devices (OLED), as a model for graphene and carbon nanoflakes in quantum-chemical studies, and as a molecule exhibiting anomalous fluorescence. In 1981, Amirav et al. \cite{amirav_1980, amirav_1981} reported the dispersed fluorescence and fluorescence excitation spectra of C$_{32}$H$_{14}$ in a supersonic jet. The observed absorption bands were assigned to the two lowest electronically excited states of C$_{32}$H$_{14}$, \textit{S}$_{1}$(\textit{B}$_{3\text{u}}$) and \textit{S}$_{2}$(\textit{B}$_{2u}$), based on differences in intensities and radiative emission lifetimes, and a zero-zero transition energy for the \textit{S}$_{1}$(\textit{B}$_{3\text{u}}$) state of 21499 $\pm$ 10 cm$^{-1}$ was inferred. In 2024, we re-recorded dispersed fluorescence and fluorescence excitation spectra of C$_{32}$H$_{14}$ isolated in solid \textit{para}-H$_{2}$ \cite{weber2024}. Based on our experimental results and spectra simulated with a Franck-Condon Herzberg-Teller model, we reassigned the spectral bands originally assigned to the \textit{S}$_{1}$--\textit{S}$_{0}$ transition by Amirav et al. \cite{amirav_1980, amirav_1981} to the \textit{S}$_{2}$--\textit{S}$_{0}$ transition. A detailed description of our experiment and the accompanying spectral analysis is provided in \cite{weber2024}.

\begin{figure}[bh]
    \centering
    \includegraphics[width=1\linewidth]{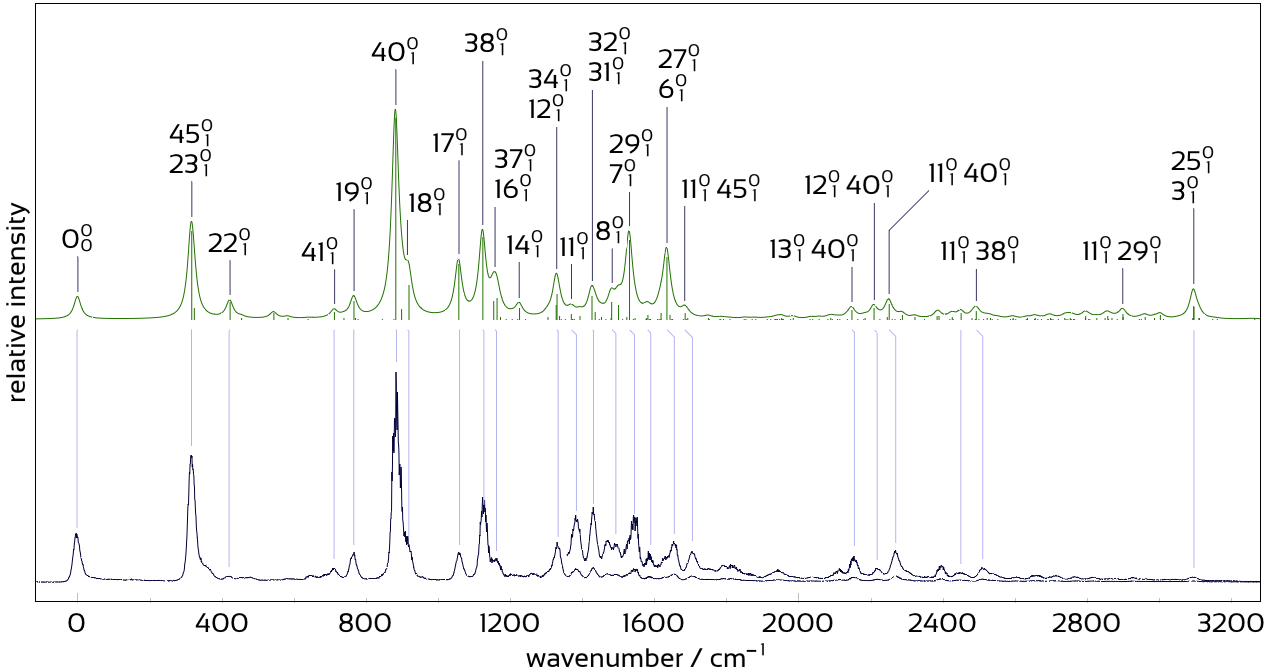}
    \caption{Comparison of the experimental emission spectrum of ovalene (bottom, with the 3x-intensity data overlaid for $\tilde{\nu}>1350$ cm$^{-1}$) with the simulated \textcolor{OliveGreen}{\textit{S}$_{2}$ $\rightarrow$ \textit{S}$_{0}$} emission spectrum (top). Individual vibronic transitions are visualized as sticks, and labeled according to user-specified settings. Vertical lines indicate the automatically matched peaks. This figure is a direct screenshot of \spec.}
    \label{fig:ovalene_emission_spectrum}
\end{figure}

Upon import of the experimental dispersed fluorescence and fluorescence excitation spectra and the corresponding computational data, \spec offers a comprehensive and interactive comparison of experiment and simulation. We compared the dispersed fluorescence spectrum of C$_{32}$H$_{14}$ isolated in solid \textit{para}-H$_{2}$ to the simulated emission spectra associated with emission from the six lowest electronically excited singlet states. From this comparison it became obvious that our experimental spectrum is best described by the simulated \textit{S}$_{2}$ $\rightarrow$ \textit{S}$_{0}$ emission spectrum and not by the \textit{S}$_{1}$ $\rightarrow$ \textit{S}$_{0}$ spectrum, prompting us to revise the assignment originally proposed by Amirav et {al.} \cite{amirav_1980, amirav_1981}. To assign the observed emission bands to individual vibronic transitions, we then aligned the simulated \textit{S}$_{2}$ $\rightarrow$ \textit{S}$_{0}$ emission spectrum with the experimental spectrum (as shown in Fig.~\ref{fig:ovalene_emission_spectrum}) by fine-tuning the parameters --- including  $\tilde{\nu}_{00}$, overall intensity, and Lorentzian FWHM -- of the convoluted stick spectrum in \specs GUI. The alignment is immediately reflected in the plot.

Most of the computed peaks align best with the experimental data when a vibrational frequency scaling factor of 0.988 is applied. The simulation also predicts contributions from fundamental C--H stretching vibrations around 3100 cm$^{-1}$. Leveraging \specs support for vibration type specific wavenumber scaling, we selectively adjusted the position of the predicted 3100 cm$^{-1}$ band by setting the scaling factor for X--H stretches to 0.977, thereby further improving the alignment of experimental and simulated data. These scaling factors are in good agreement with the literature, e.g. \cite{wheeler_2004}. The resulting table of peak assignments, exported directly from \spec, is shown in Table~\ref{tab:ovalene_emission_assingments}.

\begin{figure}[thb]
    \centering
    \includegraphics[width=1\linewidth]{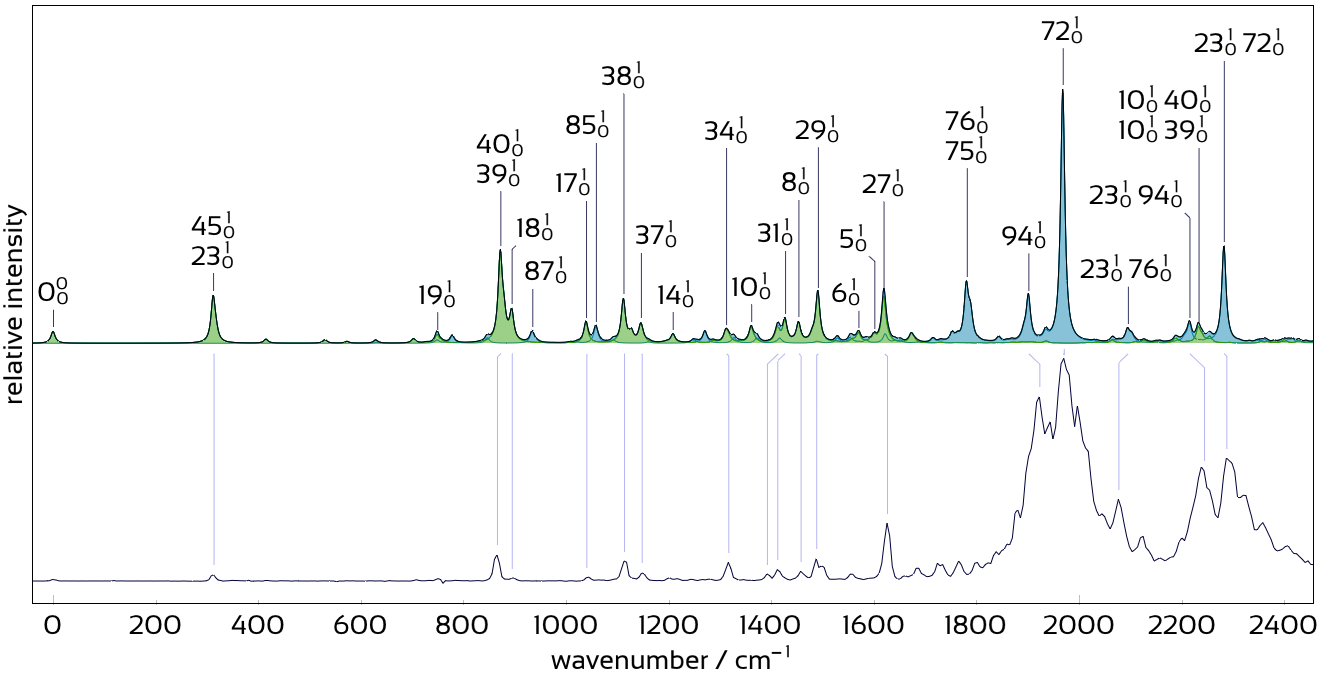}
    \caption{Experimental excitation spectrum of ovalene (bottom) and simulated excitation spectrum (top) composed of \textcolor{OliveGreen}{\textit{S}$_{2}$ $\rightarrow$ \textit{S}$_{0}$}
 and \textcolor{RoyalBlue}{\textit{S}$_{3}$ $\rightarrow$ \textit{S}$_{0}$} contributions, indicated \textit{via} shading. Peak matching was performed automatically. This figure is obtained as a screenshot of \spec.}
    \label{fig:ovalene_excitaton_spectrum}
\end{figure}

Compared to the dispersed fluorescence spectrum of C$_{32}$H$_{14}$ which can be satisfactorily fitted assuming contributions from one electronically excited state only, the fluorescence excitation spectrum of C$_{32}$H$_{14}$ is more complex and the simulated \textit{S}$_{2}$ $\leftarrow$ \textit{S}$_{0}$ spectrum alone cannot account for all observed absorption features. Overlapping the simulated \textit{S}$_{2}$ $\leftarrow$ \textit{S}$_{0}$  and \textit{S}$_{3}$ $\leftarrow$ \textit{S}$_{0}$ spectra using \specs feature to generate shaded composite spectra, as shown in Fig.~\ref{fig:ovalene_excitaton_spectrum} we were able to disentangle the contributions of these two electronic excited states to the fluorescence excitation spectrum of C$_{32}$H$_{14}$ and to infer an \textit{S}$_{3}$ - \textit{S}$_{2}$ energy gap of approximately 350 cm$^{-1}$ consistent with earlier work.

\section{Conclusion} \label{sec:conclusion}

We report here \spec, a graphical computer tool for interactive interpretation and assignment of experimental excitation and emission vibronic spectra of medium size organic molecules using a series of precomputed Franck--Condon Herzberg--Teller (FC/HT) integrals. The FC/HT data are provided to \spec as a set of Gaussian output files from the \texttt{fcht} module. 

The developed program has several distinct features making it very useful for potential users:
\begin{enumerate}
    \item The automatic analysis of the provided experimental spectra identifies distinct spectral features using user-specified pattern recognition thresholds. 
    \item The theoretical vibronic spectra for each of the precomputed excited states are displayed graphically can be interactively manipulated and superimposed with the experimental results for its convenient interpretation.
    \item \spec automatically classifies vibrational modes into types (such as X--H stretches or out-of-plane bends) and allows users to apply type-specific wavenumber scaling factors, enabling fine-tuned corrections effectively mimicking anharmonic effects.
    \item Transition labels in the plot are auto-generated upon request. Each vibrational mode can be animated by clicking its label.
    \item Whenever the interpretation of experimental spectra requires more than one excited state, the overlapping spectral patterns originating from all involved states can be easily combined into a composite theoretical spectrum matching the experimental data.
    \item Once the theoretical counterpart of the experimental spectrum has been created, \spec automatically assigns all the matching spectral features in the experimental data using user-specified similarity thresholds, producing camera-ready figures and formatted tables (see Table~\ref{tab:ovalene_emission_assingments} and Figs.~\ref{fig:ovalene_emission_spectrum} and \ref{fig:ovalene_excitaton_spectrum}) that can be directly included in publications or exported for further analysis. 
\end{enumerate}

We hope that the flexible and broad functionality of \spec will provide substantial assistance in the process of interpretation and assignment of vibronic spectra for many experimental groups.

\section*{Acknowledgments}

This work was supported by National Science and Technology Council, Taiwan (grants NSTC113-2639-M-A49-002-ASP and NSTC113-2113-M-A49-001) and Center for Emergent Functional Matter Science of National Yang Ming Chiao Tung University from The Featured Areas Research Center Program within the framework of the Higher Education Sprout Project by the Ministry of Education (MOE) in Taiwan.

\end{document}